\documentclass[letter]{aa}
\usepackage{graphicx}% Include figure files
\usepackage{dcolumn}% Align table columns on decimal point
\usepackage{bm}% bold math
\usepackage{txfonts}
\usepackage{natbib}
\usepackage{amsmath}    % need for subequations
\usepackage{graphicx}   % need for figures
\usepackage{verbatim}   % useful for program listings
\usepackage{color}      % use if color is used in text
\usepackage{subfigure}  % use for side-by-side figures
\usepackage{gensymb} %for degree symbols and others
\usepackage{wasysym} %for astro symbols
\usepackage{hyperref}   % use for hypertext links, including those to external documents and URLs
\bibpunct{(}{)}{;}{a}{}{,} % to follow the A&A style
\usepackage{graphicx}
\usepackage{nth}
\newcommand{\kms}{\mbox{${\rm km\,s}^{-1}$}}

\newcommand{\Mjup}{\mbox{${M}_{J}$}}

\newcommand{\insolation}{\mbox{${S}_{\oplus}$}}

\newcommand{\Rjup}{\mbox{${R}_{J}$}}

\begin{document}

\title{Hot Exoplanet Atmospheres Resolved 
with Transit Spectroscopy (HEARTS)\thanks{Based on observations made at ESO 3.6~m telescope (La Silla, Chile) under ESO programme 098.C-0305 (PI Ehrenreich).}}
\subtitle{V. Detection of sodium on the bloated super-Neptune WASP-166b}
\author{J.~V.~Seidel\inst{1} 
\and D.~Ehrenreich\inst{1}
\and V.~Bourrier\inst{1}
\and R.~Allart\inst{1}
\and O.~Attia\inst{1} % omar.attia@polytechnique.edu
\and H.~J.~Hoeijmakers\inst{1,2,3}
%observers
\and M.~Lendl\inst{1,4}
\and E.~Linder\inst{2} % esther.linder@space.unibe.ch
\and A.~Wyttenbach\inst{5}
%for general comments & because on proposal
\and N.~Astudillo-Defru\inst{6} %nastudillo@ucsc.cl
\and D.~Bayliss\inst{7} %d.bayliss@warwick.ac.uk
\and H.~M.~Cegla\inst{1,7}
\and Kevin~Heng\inst{2}
\and B.~Lavie\inst{1}
\and C.~Lovis\inst{1}
\and C.~Melo\inst{8} %cmelo@eso.org
\and F.~Pepe\inst{1}
\and L.~A.~dos~Santos\inst{1}
\and D.~S\'egransan \inst{1}
\and S.~Udry\inst{1}
}

\institute{Observatoire astronomique de l'Universit\'e de Gen\`eve, chemin des Maillettes 51, 1290 Versoix, Switzerland
\and University of Bern, Center for Space and Habitability, Gesellschaftsstrasse 6, CH-3012, Bern, Switzerland
\and Lund Observatory, Department of Astronomy and Theoretical Physics, Lunds Universitet, Solvegatan 9, 222 24 Lund, Sweden
\and Space Research Institute, Austrian Academy of Sciences, Schmiedlstr. 6, 8042, Graz, Austria
\and Universit\'e Grenoble Alpes, CNRS, IPAG, 38000 Grenoble, France
\and Departamento de Matem\'atica y F\'isica Aplicadas, Universidad Cat\'olica de la Sant\'isima Concepci\'on, Alonso de Rivera 2850, Concepci\'on, Chile
\and Department of Physics, University of Warwick, UK
\and European Southern Observatory, Alonso de C\'ordova 3107, Vitacura, Regi\'on Metropolitana, Chile
}

\abstract{Planet formation processes or evolution mechanisms are surmised to be at the origin of the hot Neptune desert. Studying exoplanets currently living within or at the edge of this desert could allow disentangling the respective roles of formation and evolution. We present the HARPS transmission spectrum of the bloated super-Neptune WASP-166b, located at the outer rim of the Neptune desert. Neutral sodium is detected at the $3.4\,\sigma$ level ($0.455\pm0.135\,\%$), with a tentative indication of line broadening, which could be caused by winds blowing sodium farther into space, a possible manifestation of the bloated character of these highly irradiated worlds. We put this detection into context with previous work claiming a non-detection of sodium in the same observations and show that the high noise in the trace of the discarded stellar sodium lines was responsible for the non-detection. We highlight the impact of this low signal-to-noise remnant on detections for exoplanets similar to WASP-166b.}
\keywords{Planetary Systems -- Planets and satellites: atmospheres, individual: WASP-166b -- Techniques: spectroscopic -- Instrumentation: spectrographs -- Methods: observational}
\maketitle

%----------------------------------------------------------------------------------------
%       ARTICLE CONTENTS
%----------------------------------------------------------------------------------------
\section{Introduction}

\begin{figure}[htb]
\resizebox{\columnwidth}{!}{\includegraphics[trim=3.0cm 10.0cm 3.5cm 10.5cm]{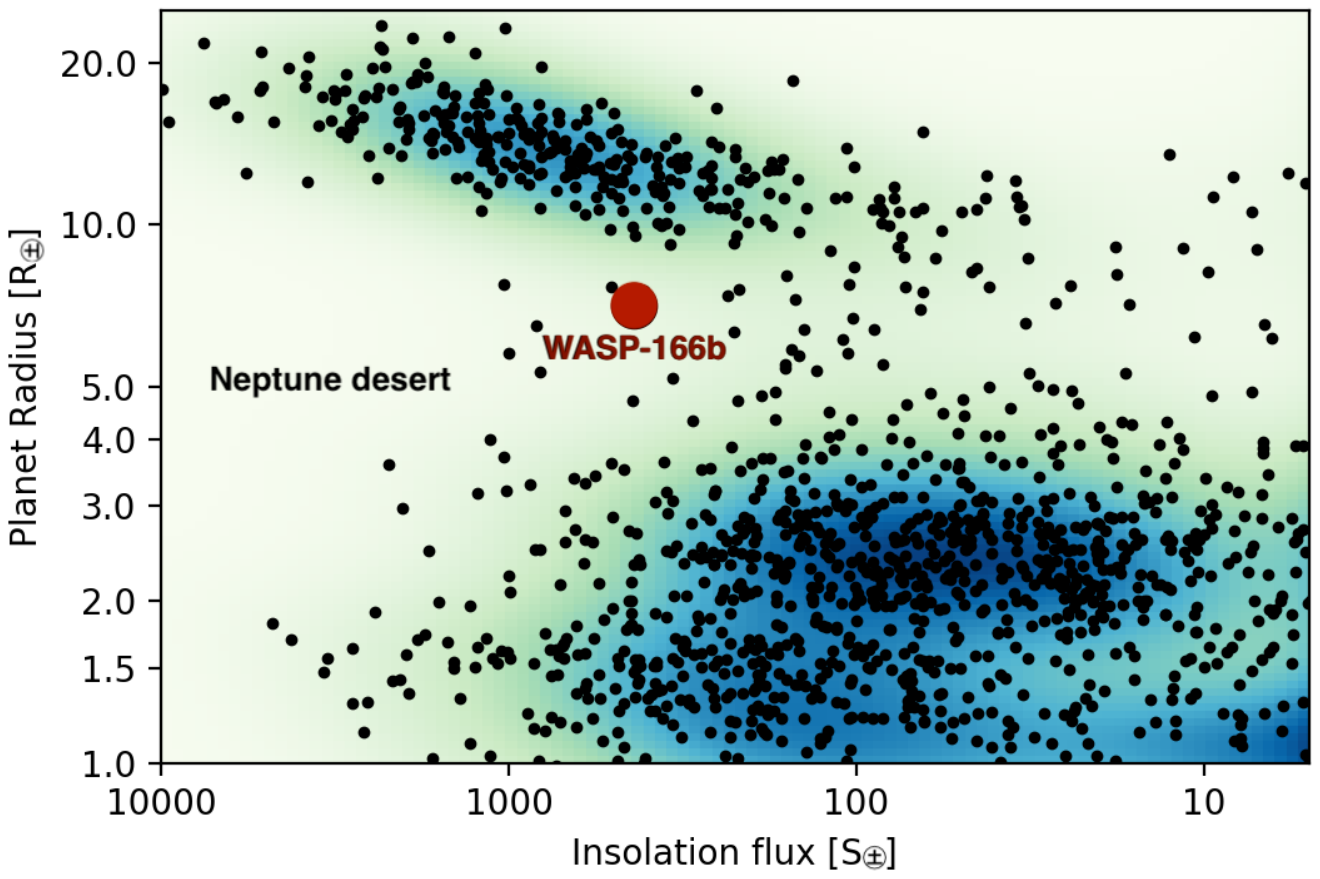}}
        \caption{Insolation vs radius diagram highlighting WASP-166b as a larger red dot in the landscape of current exoplanet detections at the edge of the Neptune desert.}
        \label{fig:massflux}
\end{figure}

One of the most prominent features of the current landscape of exoplanets is the Neptune desert, an area in the radius-insolation diagram with a lack of strongly irradiated Neptune-sized planets \citep{Lecavelier2007, Beauge2013, Mazeh2016}. This feature is not an observational bias; these worlds are accessible via various observational methods. Starting with the first detection of an evaporating atmosphere \citep{Vidal2003}, both theoretical \citep[e. g.][]{Owen2018,Owen2019} and observational \citep[e. g.][]{Ehrenreich2015, Bourrier2018} results show that atmospheric escape is a dominant process in shaping the desert. The high irradiation received by these close-in planets, both in the UV and X-ray bands, triggers a significant expansion of the upper atmosphere \citep{Lammer2003}, indicating that warm and hot Neptunes potentially erode over time \citep{Lecavelier2007,Owen2012}. In this context, WASP-166b presents a rare opportunity to study a planet within the desert, made especially interesting by its low density ($\rho = 0.54\pm0.09\, \mathrm{g}/\mathrm{cm}^3$) suggesting a bloated atmosphere \citep{Hellier2019, Bryant2020} (see Figure \ref{fig:massflux}).

\begin{figure*}[htb]
\resizebox{\textwidth}{!}{\includegraphics[trim=2.3cm 5.5cm 2.5cm 5.6cm]{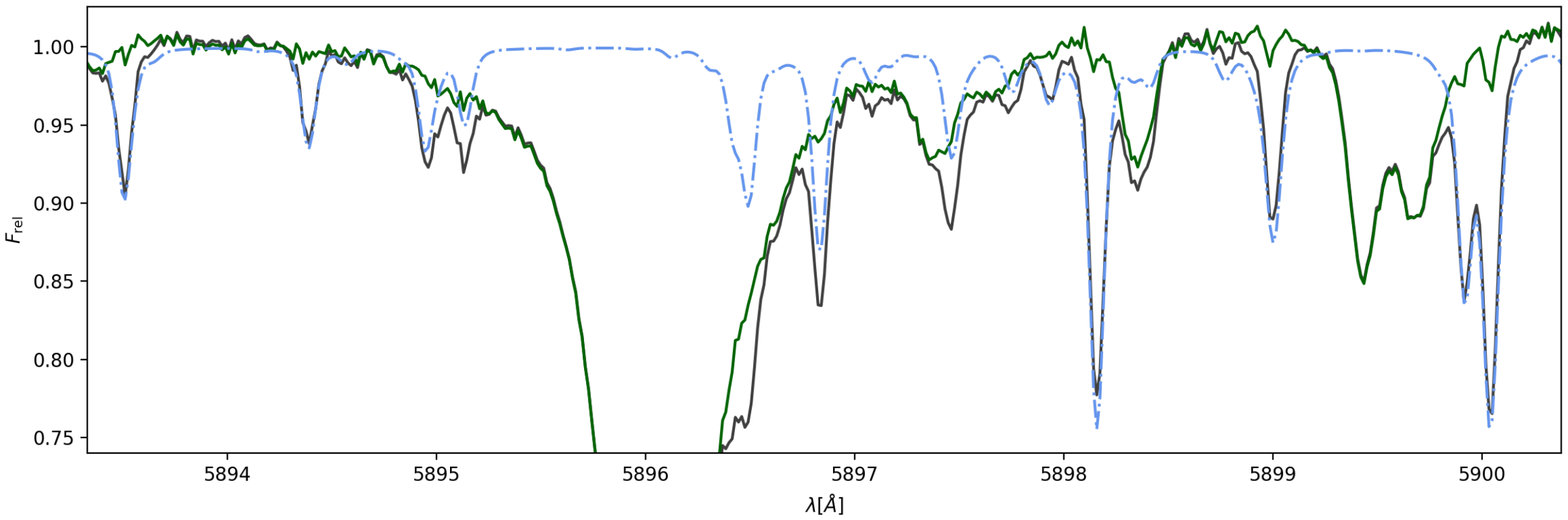}}
        \caption{The normalised sum over all spectra taken during night 1 around the sodium D1 line at $\sim 5896\,\AA$. The normalised sum before telluric correction is shown in black, after the application of the telluric correction in dark green. The light blue, dash-dotted line indicates the telluric profile generated for the spectra at transit centre for night 1. The telluric influence on the spectra was corrected down to the noise level and do not impact the transmission spectrum.}
        \label{fig:tell_check}
\end{figure*}

\noindent Neptune-sized worlds remain challenging observational targets due to their size and subsequent lower signal-to-noise ratio (SNR) compared to Jupiter-sized targets. In this work, we present the observations our group obtained as part of the HEARTS survey and highlight pitfalls in analysing low SNR data, most importantly effects from the stellar sodium lines.
 We then present our detection of neutral sodium in the atmosphere of WASP-166b and lastly, comment on the previous, independent analysis of the same observations and non-detection of sodium in \cite{Zak2019} (from here on Z2019).

\section{Observations and data reduction}
\label{sec:obs}

 \begin{table*}
\caption{Log of observations.}
\label{table:nightoverview}
\centering
\begin{tabular}{c c c c c c c c }
\hline
\hline
&Date   &$\#$Spectra \tablefootmark{a}  &Exp. Time [s]  &Airmass        \tablefootmark{b}&Seeing        & SNR \#56 & SNR stellar line core     \\
\hline
Night 1&2017-01-13&57 (26/32)&350, 300 &1.95-1.05-1.20&0.7-2.5&30 - 62 & 7-15\\
Night 2&2017-03-03&32 (27/5)\tablefootmark{c} &400, 300&1.01-1.20-2.40&0.7-1.3&30 -  74 & 7-18\\
Night 3&2017-03-14&53 (23/30)&350&1.20-1.05-2.10&0.6-1.7&40 - 68 & 9-16\\
\hline
\end{tabular}
\tablefoot{\tablefoottext{a}{Parenthesis show in- and out-of-transit spectra after excluding exposures.}
\tablefoottext{b}{Airmass at the beginning, centre, and end of each transit.}\tablefoottext{c}{This night was discarded from the analysis.}}
\end{table*}

The observations consist of three spectroscopic transits of the bloated super-Neptune WASP-166b in front of WASP-166, a bright F9 star (Vmag = 	$9.36$, distance=$113.0\pm1.0$ pc). The observations made use of the HARPS (High Accuracy Radial velocity Planet Searcher) echelle spectrograph at the ESO 3.6 m telescope in La Silla Observatory, Chile \citep{Ma03}. They were performed on the 2017-Jan-13, 2017-Mar-03, and 2017-Mar-14 as part of the HEARTS survey (ESO programme: 098.C-0305; PI: Ehrenreich). An overview of the nights can be found in Table \ref{table:nightoverview}.

\noindent Multiple transits were observed to ensure reproducibility as well as a sufficiently high SNR. The e2ds spectra were processed by the HARPS Data Reduction Pipeline (DRS v3.5). This work focusses on the sodium doublet (spectral order $56$, $5850.24$~\r{A} - $5916.17$~\r{A}). 
The data were taken during the transit (in-transit spectra), but also before and after the transit (out-of-transit spectra). A sufficient sample of out-of-transit spectra is necessary to accurately extract the planet's atmospheric signal from the total flux during transit. In night 1 (75 spectra taken), two spectra were tests to establish the correct exposure time and thus rejected. One out-of-transit spectrum was rejected due to overall low flux, thirteen in-transit spectra due to the possible influence of low-SNR remnants on the signal (see Section \ref{sec:stellar}), and one due to a passing cloud. In night 2 (52 spectra taken), twelve in-transit spectra were rejected due to low-SNR remnants, and all out-of transit spectra after the transit because of a steep fall in SNR. Two out-of-transit spectra before the transit have differing exposure times. This led to an insufficient baseline, making a correct separation of the in-transit signal from the stellar spectrum impossible. As a result, night 2 was discarded from the sample. In night 3 (65 spectra taken), two spectra were rejected due to low flux, eleven in-transit due to low-SNR remnants.

The data are corrected for the blaze, cosmic rays, and telluric absorption lines. Telluric sodium was monitored with the detector's fibre B and not found in either night. The remaining telluric lines were corrected with {\tt molecfit} version 1.5.1. \citep{Sm15, Ka15}, a well-established tool to correct telluric features in ground-based observations provided by ESO \citep[see][for applications and further details]{Al17, Seidel2019}. We controlled the telluric correction by comparing the sum over all spectra in the observer's rest frame before and after the application of {\tt molecfit}. Summing over all spectra makes any cumulative effects due to an imprecise telluric correction visible. As can be seen in Figure \ref{fig:tell_check}, the combined telluric effect was reduced to the noise level, without influencing any stellar lines. All system parameters used in this analysis were taken from \cite{Hellier2019}.

%Additionally, if any telluric sodium was nonetheless present, it would be Doppler-shifted by the barycentric Earth radial velocity in the transmission spectrum, by $\sim17\kms$ for night 1 and $\sim-8\kms$ for night 3 as well as by the , and not overlap significantly with any planetary sodium signature.
\section{Previous analysis of the observations}
\label{sec:Zak}

\begin{figure}[htb]
\resizebox{\columnwidth}{!}{\includegraphics[trim=4.0cm 9.0cm 4.0cm 9.0cm]{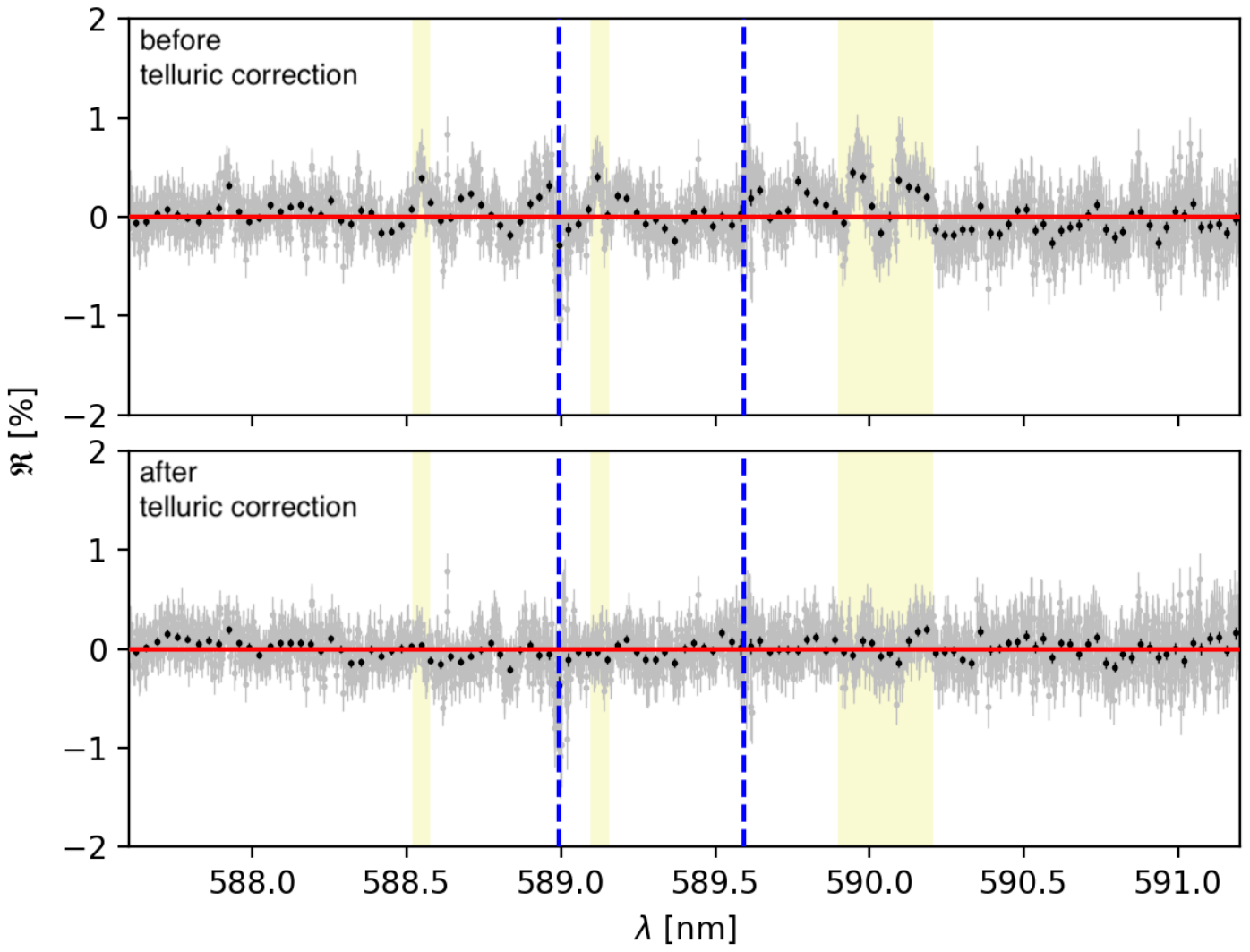}}
        \caption{Recreation of the analysis as presented in Z2019 of the same observations used in this work without telluric correction (upper panel) and with telluric correction using {\tt molecfit} (lower panel). Colours, scales and binning recreate Figure 5 of Z2019, the red line indicates the mean, the blue dashed line the expected positions of the sodium lines. Comparing Figure 5 in Z2019 with our recreation attempt shows that Z2019 has residuals from telluric lines (most pronounced in wavelength bands highlighted in light yellow).}
        \label{fig:tell_before_after}
\end{figure}

The presented observations were previously analysed in Z2019, who reported a non-detection and set an upper limit to a sodium absorption signal at $0.14\,\%$. For comparison purposes, we have tried to recreate the results presented in Z2019, which included telluric correction, but no pre-selection of spectra based on observational conditions, no correction for stellar effects nor for the RossiterMcLaughlin (RM)-effect (for further details on the RM-effect see \cite{Rossiter1924, McLaughlin1924, Lo15, Cegla2016}). However, when comparing Figure 5 of Z2019 with our recreation of their analysis (Figure \ref{fig:tell_before_after}, lower panel), the spectra show discrepancies. Our noise level is visibly lower than in the transmission spectrum presented in Z2019, albeit the use of the same spectra from the same observations. Comparing the yellow bands in Figure \ref{fig:tell_before_after} with the same wavelength range in Figure 5 of Z2019 shows a clear correlation between the cumulative effect of telluric lines and the differences between the results of Z2019 and our attempted recreation. A closer look at Figure 1 in Z2019, which depicts the telluric correction via direct calculation of a telluric spectrum, shows that the telluric lines in the mentioned wavelength ranges were not corrected down to the noise level and thus create residuals in the transmission spectrum. Setting aside the telluric contamination in Z2019, we were able to recreate their analysis. In the following, we will highlight various effects that may have masked the sodium feature in Z2019, and present a sodium detection at the $3.4\,\sigma$ level.

\section{Transmission spectroscopy of WASP-166b}
\label{sec:fulltrans}

\begin{figure*}[htb]
\resizebox{\textwidth}{!}{\includegraphics[trim=1.0cm 9.0cm 1.0cm 9.2cm]{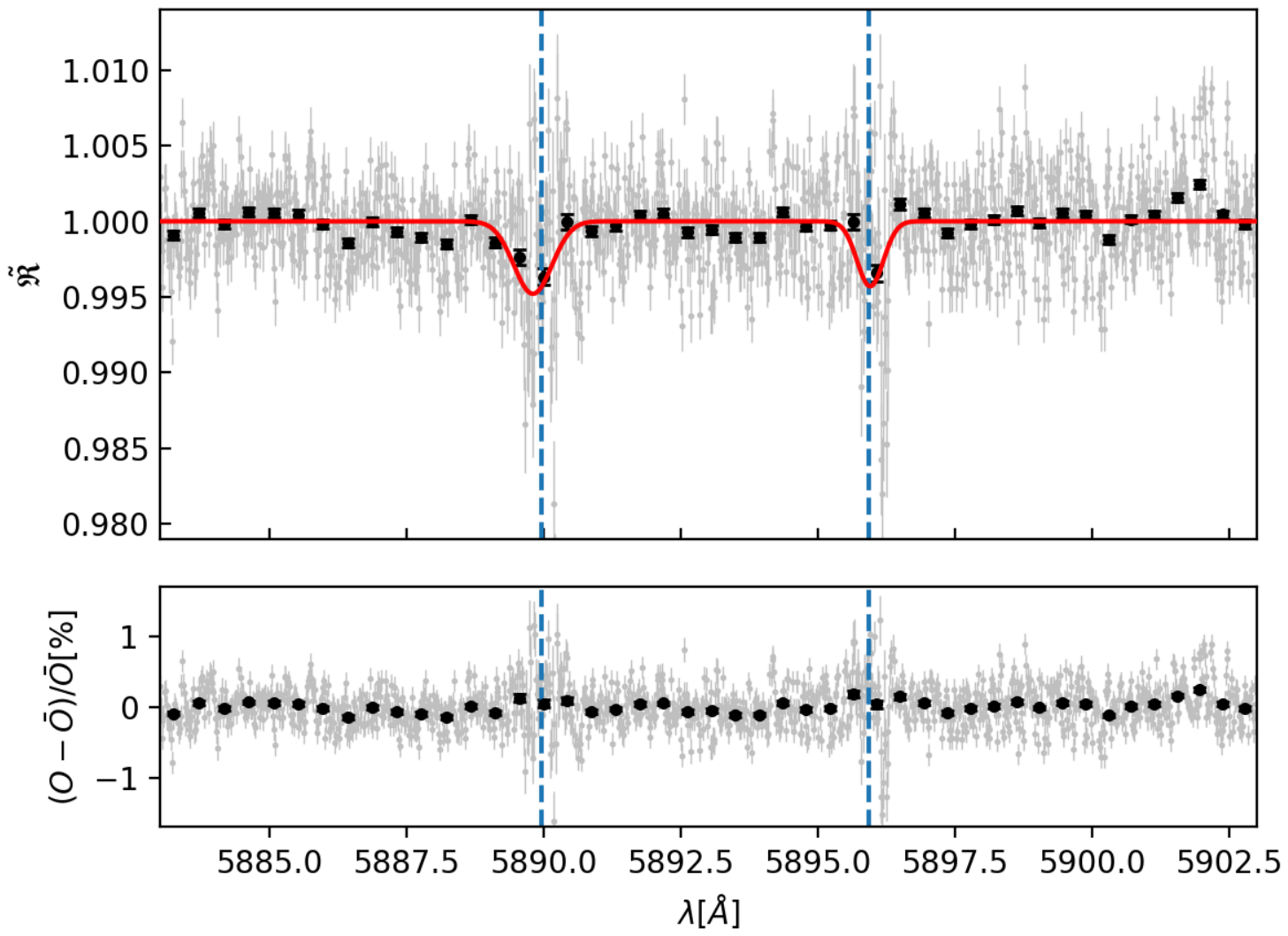}}
        \caption{HARPS sodium doublet transmission spectrum of WASP-166b for all nights combined shown in the PRF. Upper panel: In grey, data points at full HARPS resolution. In black, grey data points binned by x25 for visibility.The theoretical line centres for the sodium doublets are shown as vertical blue dashed lines, a Gaussian fit to the unbinned data is shown in red. The sodium absorption is visible for both lines of the doublet. The combined line contrast is measured at  $0.455\pm0.135\,\%$, resulting in a $3.4\,\sigma$ detection, see Sec. \ref{sec:transspec}. Lower panel: Residuals of the Gaussian fit in $\%$. }
        \label{fig:transspectrum}
\end{figure*}

We follow \cite{Wy15} in the calculation of the transmission spectrum, with the modifications in the spectra normalisation described in \cite{Seidel2019}. Compared to Z2019, we discarded night 2, and multiple in- and out-of-transit spectra (see Section \ref{sec:obs}). All spectra are shifted from the observer's rest frame into the stellar rest frame (SRF), where the planetary spectra are separated from the stellar absorption signal by dividing each in-transit spectrum by the normalised sum of the out-of-transit spectra (master-out). These in-transit spectra of the planet's atmosphere are then shifted into the planetary rest frame (PRF) and summed to create the transmission spectrum. For these shifts, we used a BERV between $17.26 - 16.62 \kms$ in night 1 and $-7.61 - -8.28 \kms$ in night 3. The system velocity is $23.62 \kms$ \citep{Hellier2019}. The planet's velocity ranges from $-23.64$ to $17.06 \kms$ in night 1 and from $-15.83$ to $25.12 \kms$ in night 3.
 The transit spectrum of WASP-166b for the two transits combined is plotted in Figure \ref{fig:transspectrum}. For the Gaussian fit (shown in red), the priors on the D1 line were restricted with the fitting parameters of the D2 line.

\subsection{Stellar effects}
\label{sec:stellar}
When not corrected, the RM-effect influences the line shape and depth of the planet's transmission spectrum, masking potential detections \citep[e.g.][]{Chen2020, CasasayasBarris2020}. 
The surface radial velocity of regions that the planet occults during transit range from $+/- \sim5\,\kms$, altering the transmission spectrum by up to $0.5\,\%$ at line centre. We applied an RM-correction to each spectrum following \cite{Wyttenbach2020}, a technique verified in \cite{Seidel2020}, however, the RM-effect is not strong enough for WASP-166b to mask the sodium signal and does not solely account for the non-detection presented in Z2019.

\noindent In the wavelength range of the stellar sodium feature, the noise is increased significantly due to the low flux (see Figures \ref{fig:2Dmap_n1} and \ref{fig:2Dmap_n3}). This effect, when shifted into the PRF, yields a transmission spectrum with low SNR in the line cores \citep{Barnes2016, Borsa2018}. In the exposures where this low-SNR remnant overlaps with the region of the planetary sodium signature, it can effectively mask any detection (likely the cause of the non-detection in Z2019). We discarded all exposures where the low-SNR remnants and the expected planetary sodium signal overlap (see Annex \ref{sec:annex} and Section \ref{sec:obs}).

\subsection{Data quality assessment}
\label{sec:bootstrap}
\begin{figure*}[htb]
\resizebox{\textwidth}{!}{\includegraphics[trim=-1.0cm 5.1cm -0.75cm 5.8cm]{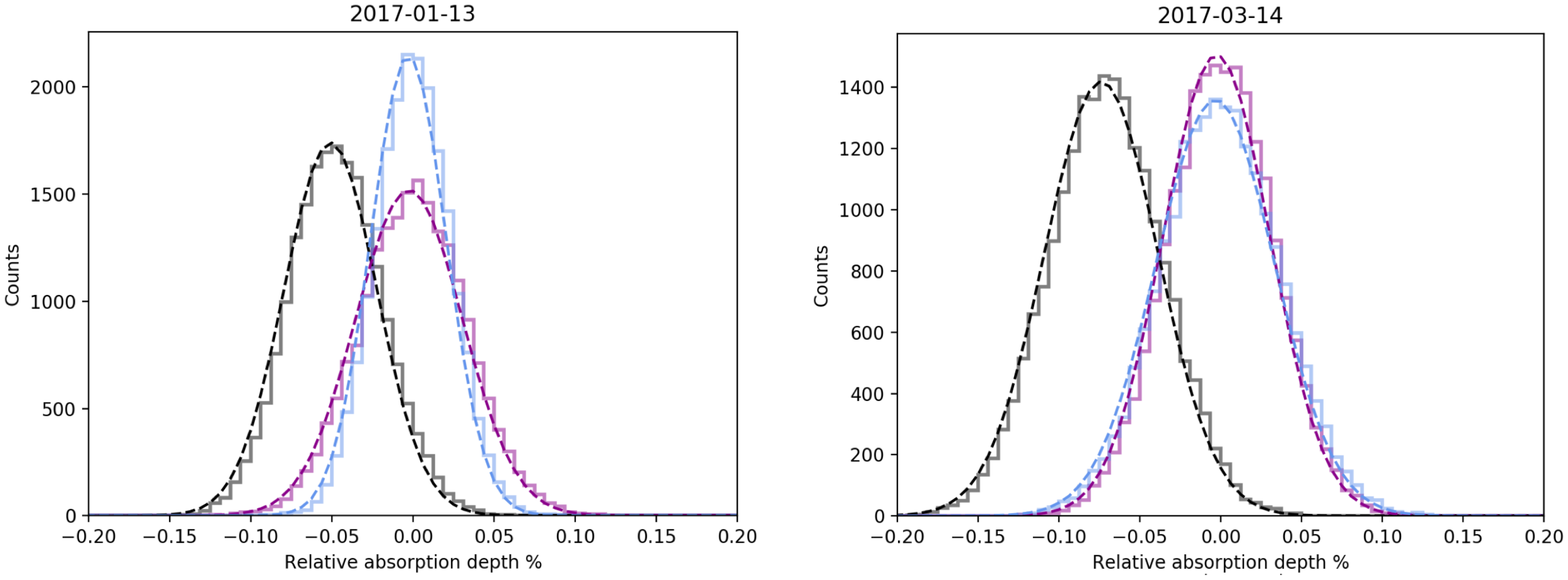}}
\caption{Distributions of the EMC bootstrap analysis for the $12$ \r{A} passband for the two transits. Both the `in-in' (purple) and `out-out' (blue) distributions are centred around zero (no sodium detection), but the randomised `in-out' distribution shows a detection (black). Each distribution was fitted with a Gaussian, shown with dashed lines. Given that each night has a different number of observed spectra, each randomisation also has a different number of counts, and thus a different scaling of the y-axis.}
\label{fig:random}
\end{figure*}

To assess the probability that a signal in our transmission spectrum is a false-positive, stemming from spurious events like instrumental effects, observational conditions, or stellar spots, we performed an empirical Monte-Carlo (EMC), or bootstrap, analysis. It predicts the likelihood of a false positive for our observations, which is taken into account in the calculation of the detection level. This additional uncertainty is especially important for noisy observations, where the likelihood of false-positive signals increases. We draw sub-samples from either the in-transit or out-of-transit spectra to create randomised transmission spectra, following \cite{Re08}. The three different scenarios are in-in (all spectra drawn from the real in-transit data and attributed to virtual in-transit and out-of-transit samples), and likewise an out-out and in-out (virtual in transit drawn from real in transit and virtual out-of-transit drawn from real out-of-transit) scenario. The generated transmission spectra should show no absorption (depth centred at 0) for the in-in and the out-out case, but an absorption feature for the in-out case \citep[see][for further details]{Wy15,Seidel2019}.

\noindent The likelihood of a false positive is calculated as the standard deviation of the out-out distribution, derived from the Gaussian fit to the distribution, multiplied by the square root of the fraction of out-of-transit spectra to the total number of spectra to account for the biased sample selection \citep{Re08,As13,Wy15}. 
The results of the bootstrapping are shown in Figure \ref{fig:random}, where each run was performed with $20,000$ iterations. The in-in and out-out scenarios are both centred as expected around 0, whereas the in-out distribution is significantly shifted towards negative and compatible values (meaning absorption) for both nights. The false positive likelihood for nights 1 and 3 is $0.022\,\%$ and $0.027\,\%$ respectively.

\subsection{Atmospheric absorption depth}
\label{sec:transspec}

Figure \ref{fig:transspectrum} shows the sodium doublet transmission spectrum of WASP-166b in the PRF. The double-peaked feature of the resonant sodium doublet is clearly visible and fitted with Gaussians. We use the fitted amplitude of the sodium lines as the absorption depth and calculate its uncertainty as the sum of the false positive likelihood (see Sec. \ref{sec:bootstrap}, \citep{Re08}), the mean of the uncertainty of the data in the reference bands from \cite{Wy15}, and the uncertainty of the Gaussian fit. This results in a sodium detection of $0.455\pm0.135\,\%$, or $3.4\,\sigma$, with the line depth for the D2 line $0.479\pm0.192\,\%$ and for the D1 line $0.431\pm0.189\,\%$. The difference in line depth between the two lines is, as expected, not statistically significant.

\section{Interpretation of the obtained sodium feature}

The same dataset was analysed in Z2019, which did not report a detection of the sodium feature presented here. Z2019 reported an upper limit for sodium of $0.14\,\%$ and infer a high cloud deck on WASP-166b, obscuring any absorption features. 

\noindent In contrast, we report the detection of an absorption feature at $0.455\pm0.135\,\%$ after carefully discarding exposures (see Section \ref{sec:obs}) and correcting for the RM-effect and low-SNR remnants.

\noindent The FWHM of the coadded sodium doublet is $0.69\pm0.24\,\AA$ ($34.91\pm12.02\,\kms$). The FWHM is calculated from the Gaussian fit (see Fig. \ref{fig:transspectrum}) and corresponds to a broadening of $17.45\pm6.01\,\kms$ in each direction from the line centre, tentatively indicating significant line broadening compared to the line spread function of HARPS \citep[$2.7\,\kms$,][]{Ma03}. The escape velocity on the surface of WASP-166b is $23.83\pm0.81\,\kms$, which means that because of the large uncertainties due to noise, the broadening is within $1\sigma$ of the escape velocity. This indicates at least part of the atmospheric sodium is transported at speeds close to what is needed to overcome the planet's gravity. Comparatively, in another planet with broadened sodium lines and high irradiation, the ultra-hot Jupiter WASP-76~b, the vast majority of sodium atoms remain at velocities below the escape velocity \citep{Seidel2019}. Whether this observation is part of a trend for planets in the Neptune desert remains to be seen.

To date, there are few planets in the mass range of WASP-166b  (M = $0.101\pm0.005\,\Mjup$, R = $0.63\pm0.03\,\Rjup$,  $\rho = 0.54\pm0.09\, \mathrm{g}/\mathrm{cm}^3$, \cite{Hellier2019}), with confirmed and well constrained values for both mass and radius, to evaluate the hypothesis of a bloated atmosphere. The closest in mass, while not residing in the Neptune desert, is GJ~143b (HD~21749b) (M = $0.071\pm0.007\,\Mjup$, R = $0.233\pm0.015\,\Rjup$, $\rho = 7.0\pm1.6\,\mathrm{g}/\mathrm{cm}^3$, \cite{Dragomir2019}), with $71\,\%$ the mass of WASP-166b. However, this planet does not show a bloated atmosphere, with 12.96 times the density of WASP-166b. Considering that GJ~143b receives an insolation flux of $\sim6\,\insolation$, compared to $\sim436\,\insolation$ for WASP-166b, the hypothesis that the bloated atmosphere of WASP-166b stems from the high irradiation from its host star and subsequent winds transporting sodium upwards seems plausible. WASP-166b, therefore, adds another observational puzzle piece to the understanding of the Neptune desert.

\section{Conclusion}
\label{sec:conclusion}

We analysed 193 spectra taken of WASP-166b, with 110 spectra remaining after a careful data quality assessment. Of these, 48 spectra were taken during two separate transits of WASP-166b. This led to a detection of neutral sodium via the Frauenhofer D-doublet at the $3.4\,\sigma$ level. We ruled out spurious signals via bootstrapping and suggest that the sodium lines are broadened, with velocities around the escape velocity. Combined with the analysis from \cite{Hellier2019}, it is indicative of a possible hydrodynamically escaping atmosphere.

\noindent The detection highlights one of the intents of the HEARTS survey: to vet potential targets for the ESPRESSO spectrograph operating at ESO's 8m VLT telescope. Additional data taken with ESPRESSO might allow to better resolve the tentatively broadened line shape with a small number of transits, thus enabling us to gain a glimpse into the upper atmosphere of a bloated super-Neptune via retrieval \citep[e.g.][]{Seidel2020, Fisher2019} and to gain access to its chemical composition \citep[e.g.][]{Hoeijmakers2019, Pino2018, Pino2020}. However, due to the system architecture, the large overlap between the low-SNR remnants and the planetary signal is a challenge and any further study of this system requires an in-depth analysis and correction of this effect.

%----------------------------------------------------------------------------------------
%       ACKNOWLEDGEMENTS
%----------------------------------------------------------------------------------------

\begin{acknowledgements}
This project has received funding from the European Research Council (ERC) under the European Union's Horizon 2020 research and innovation programme (project {\sc Four Aces}; grant agreement No. 724427).
This work has been carried out within the frame of the National Centre for Competence in Research 'PlanetS' supported by the Swiss National Science Foundation (SNSF). A.W. acknowledges the financial support of the SNSF by grant number $P400P2\_186765$. N. A.-D. acknowledges the support of FONDECYT project 3180063. We would like to thank the anonymous referee for their thorough and thoughtful comments, which have significantly improved the quality of the manuscript.
\end{acknowledgements}
%----------------------------------------------------------------------------------------
%       REFERENCE LIST
%----------------------------------------------------------------------------------------
%

\bibliographystyle{aa} % style aa.bst
\bibliography{HEARTSV_WASP166b}

\begin{appendix}

\onecolumn
\section{Remnants of the stellar spectrum}
\label{sec:annex}

\begin{figure*}[htb]
\resizebox{\textwidth}{!}{\includegraphics[trim=2.0cm 9.0cm 2.0cm 9.0cm]{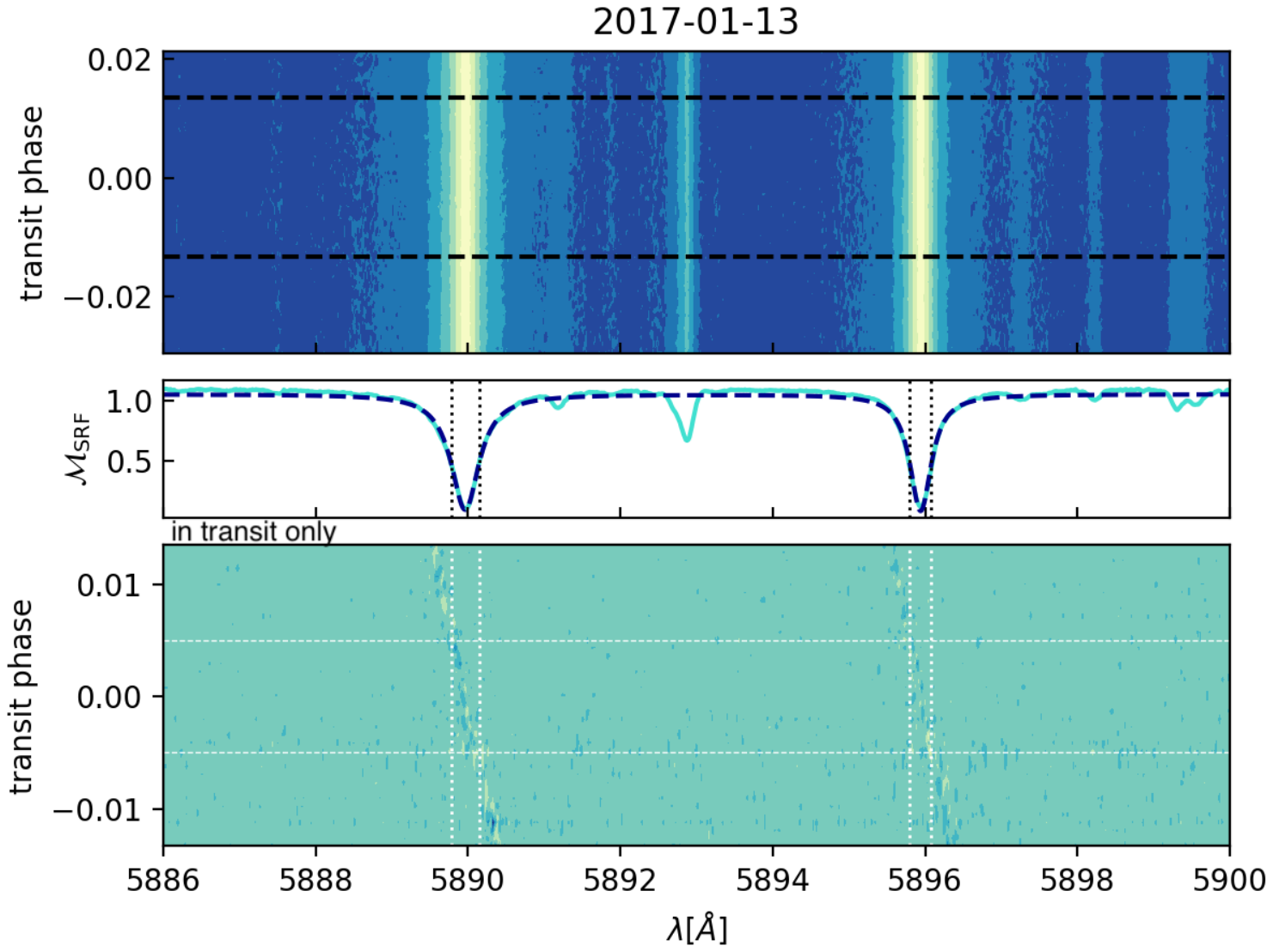}}
        \caption{The upper panel shows all spectra in the SRF as a 2D map of wavelength and transit phase for the first transit. The stellar sodium doublet is visible as two horizontal light yellow bands. Transit ingress and egress are marked with black dashed lines. The central plot shows the normalised sum of all spectra with a fit to each line in dashed blue. The FWHM is indicated as dotted vertical lines. The lower panel shows the same data, but corrected for the stellar spectrum by the master-out, in the PRF. The dotted lines propagate the position of the FWHM from the central panel. The low-SNR remnants are clearly visible, but the SNR is too low to see the planetary trace.}
        \label{fig:2Dmap_n1}
\end{figure*}

\begin{figure*}[htb]
\resizebox{\textwidth}{!}{\includegraphics[trim=2.0cm 9.0cm 2.0cm 9.0cm]{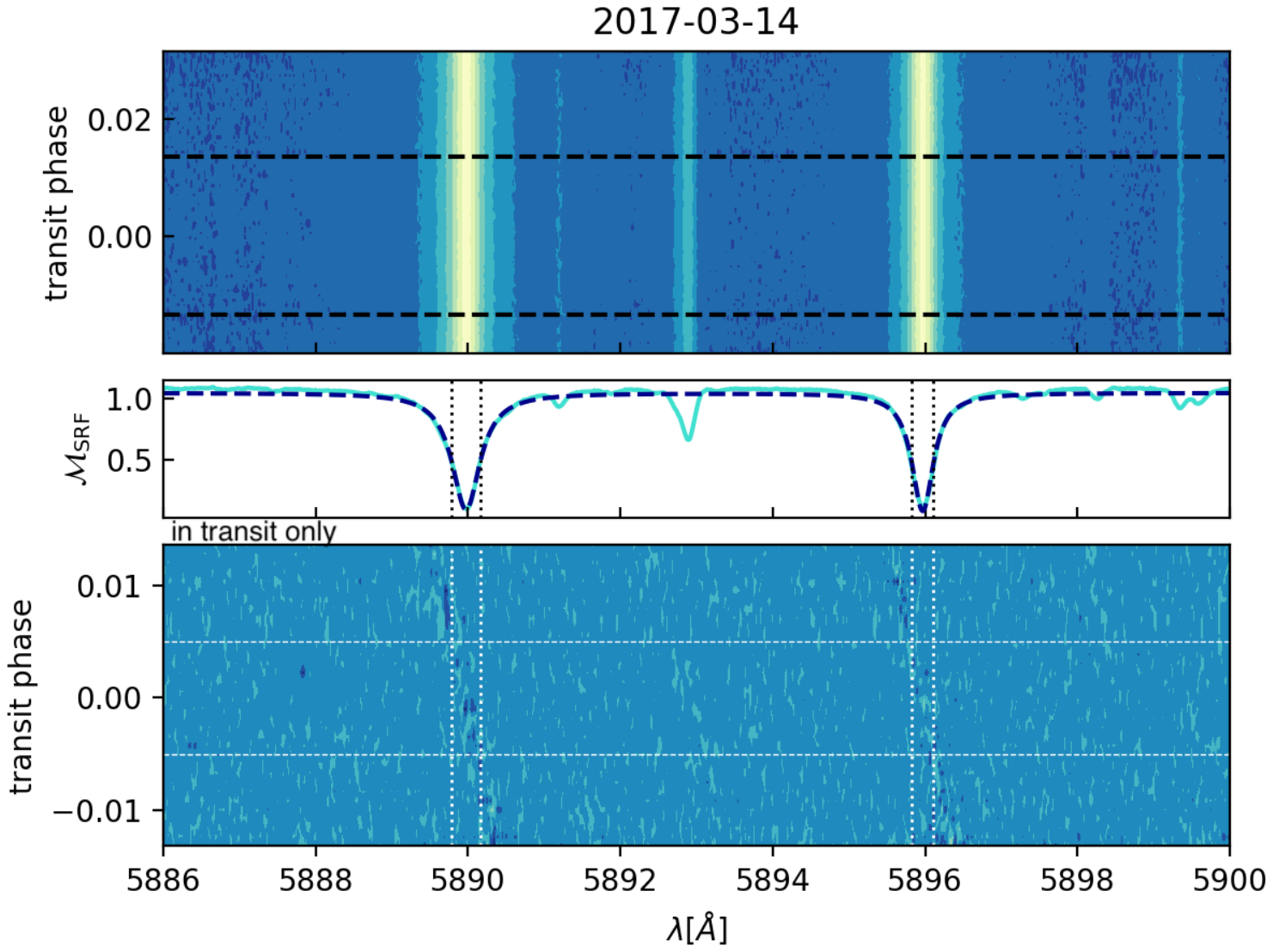}}
        \caption{Stellar remnants after correction for night 3. Please see Figure \ref{fig:2Dmap_n1} for further details on the plot. Note the general higher noise when compared to night 1. The low-SNR remnants are again clearly visible.}
        \label{fig:2Dmap_n3}
\end{figure*}

As described in Sec. \ref{sec:stellar}, the low-SNR in the bins at the centre of the stellar sodium make a correct extraction of the planetary signal challenging. We have opted for a more conservative approach in mitigating the adverse effect of the low-SNR remnants on the planetary signal. In the upper panels of Figure \ref{fig:2Dmap_n1} and \ref{fig:2Dmap_n3}, all spectra are shown in the SRF before correction. The deep features are the stellar sodium lines, with the planetary signal distributed over a broad wavelength range due to the Doppler-shift between the PRF and the SRF. Calculating the FWHM of this feature gives a maximal wavelength region for the potential planetary sodium, if it can be detected. If we now divide by the master-out and shift the remaining in-transit spectra to the PRF, the low-SNR remnant becomes visible. The slant indicates that this feature does not stem from the planet's atmosphere, but from low-SNR in the stellar sodium line core and subsequent noise in the extracted planetary spectra. For less noisy spectra, these remnants are negligible, given that they are distributed over a wide wavelength range when summing over all in-transit spectra to calculate the transmission spectrum. However, in this case, the remnants are large and exacerbated by the low change of the orbital velocity and the aligned orbit of the planet, which creates a significant overlap between the position of a potential planetary sodium signal and the low-SNR remnants. In the sum of the transmission spectrum, the remnants increase the noise significantly and can effectively hide any planetary sodium signature, or worse, produce a false positive detection of sodium. To eliminate these two possibilities, we identified all spectra where the core of the low-SNR remnants coincide with the identified maximal wavelength region for the sodium feature (white dotted lines in the lower panels), which resulted in the rejection of all spectra taken between phases $-0.005$ and $0.005$ (indicated by horizontal white dashed lines).

\end{appendix}
\end{document}